# A balanced Memristor-CMOS ternary logic family and its application

Xiao-Yuan Wang [1], *Senior Member, IEEE*, Jia-Wei Zhou [1], Chuan-Tao Dong[1], Xin-Hui Chen[1], Sanjoy Kumar Nandi[2], Robert G. Elliman[2], Sung-Mo Kang[4], *Life Fellow*, and Herbert Ho-Ching Iu[3], *Senior Member, IEEE*

**Abstract:** The design of balanced ternary digital logic circuits based on memristors and conventional CMOS devices is proposed. First, balanced ternary minimum gate TMIN, maximum gate TMAX and ternary inverters are systematically designed and verified by simulation, and then logic circuits such as ternary encoders, decoders and multiplexers are designed on this basis. Two different schemes are then used to realize the design of functional combinational logic circuits such as a balanced ternary half adder, multiplier, and numerical comparator. Finally, we report a series of comparisons and analyses of the two design schemes, which provide a reference for subsequent research and development of three-valued logic circuits.

## 1.Introduction

In the past few decades, binary logic has been widely used in digital logic systems, and large-scale integration has seen an ongoing reduction in cost. However, as the scaling speed of traditional CMOS devices gradually slows, achieving the predictions of Moore's Law is increasingly difficult[1]. Therefore, in the post-Moore era, researchers are constantly seeking new solutions, such as the use of new materials, new devices and new computing paradigms[2]-[4].

The fabrication of novel devices has enabled the continuous improvement of digital logic circuits and has provided new opportunities for the integrated circuit industry. For example, carbon nanotube field effect transistors (CNTFET) are a promising ideal transistor logic device that has larger switching current, better subthreshold characteristics and stability than CMOS devices[5][6]. However, in CNTFET-based digital logic circuits, a "charge build-up" problem occurs that may affect device switching performance[7]. In 2008, the successful fabrication of a memristor[8] also attracted the attention of researchers in various fields. It has the characteristics of nanometer size, low power consumption, and compatibility with traditional CMOS technology, and has shown great potential in the field of digital logic circuit research[9]-[11]. Notably, the non-volatility of memristors enables memory-computing integration that can realize novel computer architectures that differ from the von Neumann architecture[12][13].

With the emergence of novel logic devices, the study of multi-valued logic circuits has also attracted the attention of researchers. The multi-valued logic circuit refers to logic systems with more than 2 logic states. To achieve the same function, multi-valued logic circuits require fewer interconnections than binary logic circuit, and are expected to improve the power consumption and data density of the circuit[14][15]. As a typical multi-valued logic, ternary logic has been the most widely studied. Ternary logic representation methods are mainly divided into two categories: unbalanced and balanced. Unbalanced ternary logic includes positive {0, 1, 2} and negative {0, -1, -2} ternary states, while balanced ternary logic is represented by {-1, 0, 1} states[16]. In [17], CNTFET-based basic ternary logic gates are proposed to reduce the power consumption of the circuit compared to the CMOS-based design. In 2020, [18] proposed a design method of ternary logic gates with the resistances of the tri-valued memristor as logic variables. Subsequently, [19] proposed a ternary logic family employing a memristor-CMOS hybrid design. This used voltage as the logic variable and achieved an impressive improvement in data density.

The above studies all employed positive ternary logic, whereas this study focusses on balanced ternary logic. Compared to unbalanced ternary logic, balanced ternary logic can represent the entire range of integers without setting the sign bit[20]. In addition, its multiplication operation does not generate carry bits and the addition operation only generates carry bits in 2 of the 9 input cases. Therefore, balanced ternary logic offers advantages over unbalanced ternary logic in arithmetic operations[21]. Since the memristor is compatible with CMOS technology and is smaller than conventional CMOS devices, it is more conducive to further improvements in circuit integration. Therefore, the study of memristive balanced ternary logic circuits is expected to improve information storage density and information transmission efficiency.

The rest of this paper is organized as follows: Section 2 introduces the balanced ternary basic logic gates, including ternary minimum gate TMIN, maximum gate TMAX, and three forms of inverter. In the third section, we design balanced ternary encoder circuits, including a 3-1 encoder and a 9-2 encoder. Section 4 first introduces the design of a balanced ternary 1-3 decoder and 2-9 decoder, and then further designs functional combinational logic circuits based on the decoder, including a half adder, multiplier and numerical comparator. In Section 5, a novel balanced ternary multiplexer circuit is introduced and combinational logic circuits, including a half adder, multiplier, and numerical comparator are constructed based on this multiplexer. The circuits designed in each section are simulated and verified by LTSpice to prove the effectiveness of the designed circuit structure. Section 6 presents the characteristics of the fabricated memristor and experimentally validates the balanced ternary half adder based on the memristor. Section 7 analyses and compares the characteristics of selected components, and investigates the power consumption of the combinational logic circuits designed by the two methods. Finally, Section 8 summarizes the conclusions of this study.

## 2. Basic Logic Gates of Balanced Ternary

In our previous work[19], we investigated unbalanced positive ternary logic {0, 1, 2} circuits. Here, we focus on balanced ternary logic {-1, 0, 1}, where logic values '-1', '0', '1' correspond to voltage levels -$V_{DD}$, 0, $V_{DD}$, respectively.

### 2.1 TMIN and TMAX Gates

Table 1 The truth table of TMIN and TMAX

| $V_{in1}$ | $V_{in2}$ | TMIN | TMAX |
|---|---|---|---|
| -1 | -1 | -1 | -1 |
| -1 | 0 | -1 | 0 |
| -1 | 1 | -1 | 1 |
| 0 | -1 | -1 | 0 |
| 0 | 0 | 0 | 0 |
| 0 | 1 | 0 | 1 |
| 1 | -1 | -1 | 1 |
| 1 | 0 | 0 | 1 |
| 1 | 1 | 1 | 1 |

Table 1 shows the truth table of the designed balanced ternary minimum (TMIN) gate and maximum (TMAX) gate circuits. As can be seen from Table 1, the outputs of the TMIN and TMAX are the operations of taking the minimum value and the maximum value of the input signals, respectively. Therefore, since its function is the same as that of the AND gate and the OR gate in the positive ternary logic, the circuit structure of Figure 1 in [19] can be used. Its working principle will not be analyzed in detail here. Figure 1 presents LTSpice simulation results for all 9 possible inputs for TMIN and TMAX using the KNOWM memristor model[22]. The value of $V_{DD}$ is 1V. It is worth noting that the above two-input gates, TMIN and TMAX, can be extended to multiple inputs by simply increasing the number of memristors. For example, for the three-input TMIN and TMAX, it is only necessary to increase the number of memristors to three, and the circuit structure can be found in the circuit design of Figure 8 in [19].

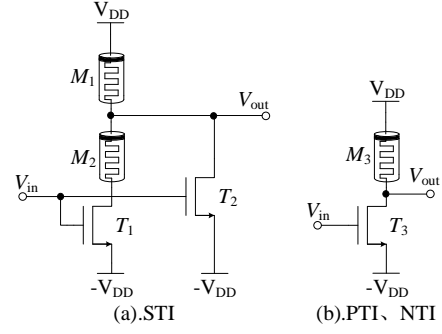

(a).STI      (b).PTI、NTI

Figure 2 The structure of balanced ternary inverters

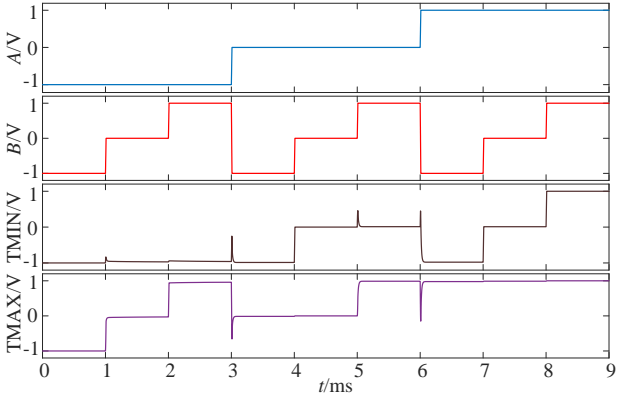

Figure 1 LTSpice simulation results of TMIN and TMAX

## 2.2 Balanced Ternary Inverters

Balanced ternary inverters also have three forms, namely Standard Ternary Inverter (STI), Positive Ternary Inverter (PTI) and Negative Ternary Inverter (NTI), and the corresponding truth table is shown in Table 2.

Table 2 Truth table of balanced ternary inverters

| $A$ | STI | PTI | NTI |
| --- | --- | --- | --- |
| -1 | 1 | 1 | 1 |
| 0 | 0 | 1 | -1 |
| 1 | -1 | -1 | -1 |

By improving our previous work[19], a novel circuit structure for the STI is obtained, as shown in Figure 2(a). By changing the connection method of the components, more precise logic operations can be achieved, and the output logic level is closer to the corresponding preset voltage value. The thresholds of the transistors $T_1$ and $T_2$ satisfy $0<v_{th1}<V_{DD}$ and $V_{DD}<v_{th2}<2V_{DD}$ respectively.

The circuit structure of the PTI and NTI gates is the same, which can be realized by changing the ground terminals of the PTI and NTI circuits in [19] to a voltage source -$V_{DD}$, as shown in Figure 2(b). Among them, the threshold voltage of $T_3$ should satisfy $V_{DD}<v_{th3}<2V_{DD}$ and $0<v_{th3}<V_{DD}$ in the PTI and NTI respectively.

Figure 3 shows the simulation results of LTSpice with three inverters. Where $v_{th1}$ takes 0.8V, $v_{th2}$ takes 1.5V, and the threshold voltages of NMOS in PTI and NTI take 1.5V and 0.8V respectively.

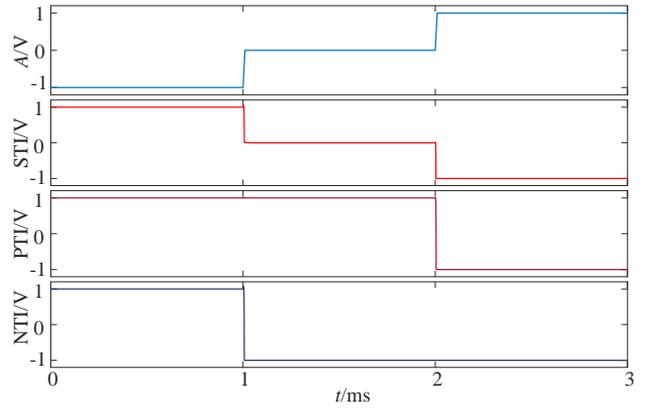

Figure 3 The simulation results of balanced ternary inverters

## 3. Encoders of Balanced Ternary Logic

In digital logic circuits, encoders are often used to distinguish a series of different things, each of which is represented by a balanced ternary code, which is what encoding means. In this section, we first design a balanced ternary 3-1 encoder and then extend it to more inputs so a balanced ternary 9-2 encoder circuit is realized.

## 3.1 Balanced Ternary 3-1 Encoder

The function of the 3-1 encoder is to encode three input signals with only high and low states (logic "1" and logic "-1") into one balanced ternary signal (logic "-1", logic "0" and logic "1"). Table 3 is the truth table of the 3-1 encoder. It is worth noting that only one encoded signal (high level signal) is allowed to be input at any time, otherwise the output will be ill-defined.

Table 3 The truth table of balanced ternary 3-1 encoder

| $X_1$ | $X_0$ | $X_{-1}$ | $Y$ |
|---|---|---|---|
| -1 | -1 | 1 | -1 |
| -1 | 1 | -1 | 0 |
| 1 | -1 | -1 | 1 |

Figure 4 is the circuit structure of the proposed balanced ternary 3-1 encoder, where the threshold voltage of the NMOS transistor $T_1$ of the Subcircuit1 module in Figure 4(b) satisfies $V_{DD}<v_{th1}<2V_{DD}$.

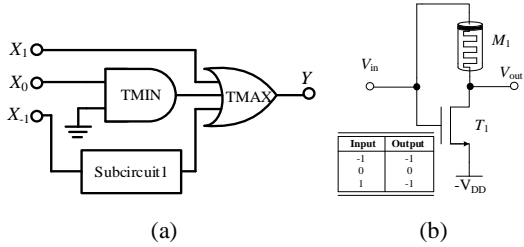

(a) (b)

Figure 4 (a). The circuit structure of balanced ternary 3-1 encoder (b). The circuit structure of Subcircuit1

When $X_1$, $X_0$ and $X_{-1}$ are "-1", "-1" and "1", respectively, the TMIN in Figure 4(a) outputs logic "-1", and the Subcircuit1 also outputs logic "-1". Therefore, the final output of the circuit is a logic "-1".

When $X_1$, $X_0$ and $X_{-1}$ are "-1", "1" and "-1", the output of the TMIN is logic "0", and the output of the Subcircuit1 is "-1". Therefore, after the maximum operation, the final output $Y$ of the circuit is logic "0".

When $X_1$, $X_0$, and $X_{-1}$ are "1", "-1" and "-1" respectively, because $X_1$ is directly connected to one of the input terminals of the TMAX, the final output of the circuit is logic "1"。

Figure 5 shows the LTSpice simulation results of the balanced ternary 3-1 encoder.

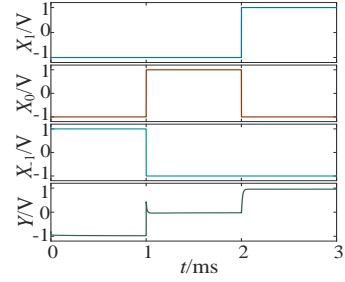

Figure 5 The simulation results of balanced ternary 3-1 encoder

## 3.2 Balanced Ternary 9-2 Encoder

As for the 3-1 encoder, the balanced ternary 9-2 encoder is also a common encoder. Its inputs are nine channels of high- or low-level signals, and the output is a 2-bit balanced ternary code. The relationship between inputs and outputs is shown in Table 4.

Table 4 The truth table of balanced ternary 9-2 encoder

| $X_4$ | $X_3$ | $X_2$ | $X_1$ | $X_0$ | $X_{-1}$ | $X_{-2}$ | $X_{-3}$ | $X_{-4}$ | $Y_1$ | $Y_0$ |
|---|---|---|---|---|---|---|---|---|---|---|
| -1 | -1 | -1 | -1 | -1 | -1 | -1 | -1 | 1 | -1 | -1 |
| -1 | -1 | -1 | -1 | -1 | -1 | -1 | 1 | -1 | -1 | 0 |
| -1 | -1 | -1 | -1 | -1 | -1 | 1 | -1 | -1 | -1 | 1 |
| -1 | -1 | -1 | -1 | -1 | 1 | -1 | -1 | -1 | 0 | -1 |
| -1 | -1 | -1 | -1 | 1 | -1 | -1 | -1 | -1 | 0 | 0 |
| -1 | -1 | -1 | 1 | -1 | -1 | -1 | -1 | -1 | 0 | 1 |
| -1 | -1 | 1 | -1 | -1 | -1 | -1 | -1 | -1 | 1 | -1 |
| -1 | 1 | -1 | -1 | -1 | -1 | -1 | -1 | -1 | 1 | 0 |
| 1 | -1 | -1 | -1 | -1 | -1 | -1 | -1 | -1 | 1 | 1 |

The circuit structure to achieve the logic in Table 4 is shown in Figure 6. The threshold voltage of $T_1$ of the Subcircuit2 in Fig. 6(b) satisfies $0<v_{th1}<V_{DD}$.

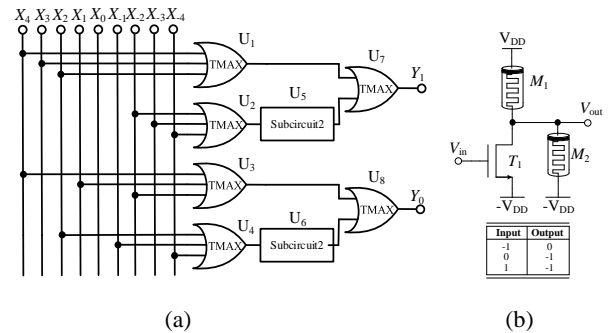

(a) (b)

Figure 6 (a). The circuit structure of balanced ternary 9-2 encoder (b). The circuit structure of Subcircuit2

For the $Y_1$ output circuit, when any one of $X_4$, $X_3$, $X_2$ is at high level, $U_1$ outputs logic "1", at this time $U_2$ outputs logic "-1" and then converted to logic "0" after Subcircuit2, so $U_7$ finally outputs logic "1". When any one of $X_1$, $X_0$, $X_{-1}$ is high, $U_1$ outputs logic "-1", $U_5$ still outputs logic "0", so $Y_1$ is logic "0". When any one of $X_{-2}$, $X_{-3}$, $X_{-4}$ is high, $U_1$ outputs logic "-1", $U_2$ is logic "1", and then converted to logic "-1" after Subcircuit2. So $U_7$ finally outputs logic "-1". The working principle of the $Y_0$ part is similar to this and will not be analyzed in detail here. Figure 7 shows the LTSpice simulation waveforms of the balanced ternary 9-2 encoder circuit.

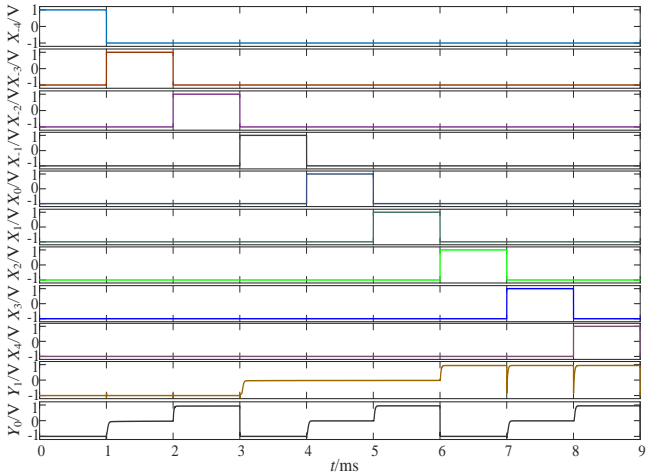

Figure 7 The simulation results of balanced ternary 9-2 encoder

## 4. Design of the Balanced Ternary Combinational Logic Circuits Based on Decoder

In this section, a balanced ternary 1-3 decoder circuit is first designed and then used to inform the design of a 2-9 decoder. In Section 4.3, we apply the decoder to the design of functional balanced ternary combinational logic circuits, including half adder, multiplier, and comparator circuits.

### 4.1 Balanced Ternary 1-3 Decoder

The function of the balanced ternary 1-3 decoder is to convert a balanced ternary signal into corresponding output high- and low-level signals. The detailed logical relationship is shown in Table 5.

Table 5 The truth table of balanced ternary 1-3 decoder

| Y | $Y_{-1}$ | $Y_0$ | $Y_1$ |
|---|---|---|---|
| -1 | 1 | -1 | -1 |
| 0 | -1 | 1 | -1 |
| 1 | -1 | -1 | 1 |

The balanced ternary 1-3 decoder can be implemented by changing all ground terminals in the positive ternary decoder of Figure 14 in [19] to $-V_{DD}$ (with 5 transistors and 7 memristors), but in order to simplify the circuit structure, the circuit is improved in this paper, and a novel balanced ternary 1-3 decoder structure (including 5 transistors and 5 memristors) is proposed. It can be seen that the components required by the circuit are reduced from 12 to 10, which reduces the circuit area. The circuit structure is shown in Figure 8, in which $Y_{-1}$ is composed of an NTI, and $Y_1$ is composed of a PTI and an NTI. The thresholds of transistors $T_2$ and $T_3$ in the $Y_0$ part both satisfy $V_{DD} < v_{th2}$, $v_{th3} < 2V_{DD}$. The working principle of the $Y_0$ part is analyzed as follows：

- **Input logic "-1":** When the input is $-V_{DD}$ (logic "-1"), $T_2$ is turned off, $T_3$ is turned on, and the memristors $M_1$ and $M_2$ form a TMIN, and the inputs are equivalent to logic "-1" and "1", so the output is logic "-1".

- **Input logic "0":** When the input is GND (logic "0"), $T_2$ and $T_3$ are both turned off, and the output terminal is pulled up to $V_{DD}$ through the memristor $M_2$, that is, the output is logic "1".

- **Input logic "1":** When the input is $V_{DD}$ (logic "1"), both $T_2$ and $T_3$ are turned on, and the output terminal will be directly connected to $-V_{DD}$ through $T_3$ and $T_2$, the output is logic "-1".

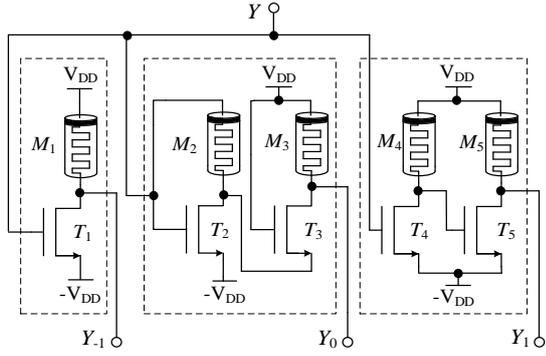

Figure 8 Novel structure of balanced ternary 1-3 decoder

Figure 9 shows the simulated waveform of the proposed balanced ternary 1-3 decoder from LTSpice, where the threshold voltages $v_{th1}$ and $v_{th5}$ of $T_1$ and $T_5$ are 0.8V, and the threshold voltages $v_{th2}$, $v_{th3}$ and $v_{th4}$ of $T_2$, $T_3$ and $T_4$ take 1.5V.

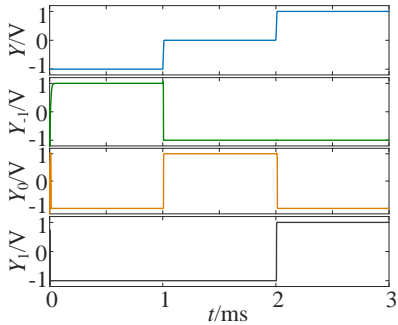

Figure 9 Simulation results of balanced ternary 1-3 decoder

### 4.2 Balanced ternary 2-9 decoder

A balanced ternary 2-9 decoder is designed based on the 1-3 decoder. The input is a two-bit balanced ternary code, and the outputs are 9 signals, which are high or low levels corresponding to the input codes. The truth table is shown in Table 6, where $A$ and $B$ are input ternary signals, and $Y_{-4}$-$Y_4$ are output signals.

Table 6 The truth table of balanced ternary 2-9 decoder

| A | B | $Y_{-4}$ | $Y_{-3}$ | $Y_{-2}$ | $Y_{-1}$ | $Y_0$ | $Y_1$ | $Y_2$ | $Y_3$ | $Y_4$ |
|---|---|---|---|---|---|---|---|---|---|---|
| -1 | -1 | 1 | -1 | -1 | -1 | -1 | -1 | -1 | -1 | -1 |
| -1 | 0 | -1 | 1 | -1 | -1 | -1 | -1 | -1 | -1 | -1 |
| -1 | 1 | -1 | -1 | 1 | -1 | -1 | -1 | -1 | -1 | -1 |
| 0 | -1 | -1 | -1 | -1 | 1 | -1 | -1 | -1 | -1 | -1 |
| 0 | 0 | -1 | -1 | -1 | -1 | 1 | -1 | -1 | -1 | -1 |
| 0 | 1 | -1 | -1 | -1 | -1 | -1 | 1 | -1 | -1 | -1 |
| 1 | -1 | -1 | -1 | -1 | -1 | -1 | -1 | 1 | -1 | -1 |
| 1 | 0 | -1 | -1 | -1 | -1 | -1 | -1 | -1 | 1 | -1 |
| 1 | 1 | -1 | -1 | -1 | -1 | -1 | -1 | -1 | -1 | 1 |

| A\B | -1 | 0 | 1 |
|---|---|---|---|
| -1 | 1 | -1 | -1 |
| 0 | -1 | -1 | -1 |
| 1 | -1 | -1 | -1 |

$Y_{-4}$

Figure 10 Karnaugh map of output terminal $Y_{-4}$ of balanced ternary 2-9 decoder

From Table 6, we can see that for each input combination, only one output is high (logic "1"), and the rest of the outputs are low (logic "-1"). Taking $Y_{-4}$ as an example, the Karnaugh map can be obtained as shown in Figure 10, and then the input-output relationship can be obtained as $Y_{-4}=A_{-1}\cdot B_{-1}$. Similarly, all the relationships between input and output can be obtained as follows:

$$\begin{cases} Y_{-4} = A_{-1}\cdot B_{-1} & Y_{-1} = A_0\cdot B_{-1} & Y_2 = A_1\cdot B_{-1} \\ Y_{-3} = A_{-1}\cdot B_0 & Y_0 = A_0\cdot B_0 & Y_3 = A_1\cdot B_0 \\ Y_{-2} = A_{-1}\cdot B_1 & Y_1 = A_0\cdot B_1 & Y_4 = A_1\cdot B_1 \end{cases} \quad (1)$$

the subscripts of $A$ and $B$ represent their values respectively, and "•" represents the minimum operation.

According to the analysis in the previous section, the balanced ternary 1-3 decoder has the characteristic that when the input is $X=k$, the corresponding output is $X_k=1$, and the other outputs are -1. Therefore, according to formula (1), we can use $A_m$, $B_n$ (m, n are -1, 0, 1) as the inputs of the TMIN, and $Y_{-4}$-$Y_4$ are obtained as the output of the TMINs. The circuit structure is shown in Figure 11. Among them, $A$ is the high-order input of the 2-9 decoder, $B$ is the low-order input, and $Y_{-4}$-$Y_4$ are the outputs of the designed balanced ternary 2-9 decoder.

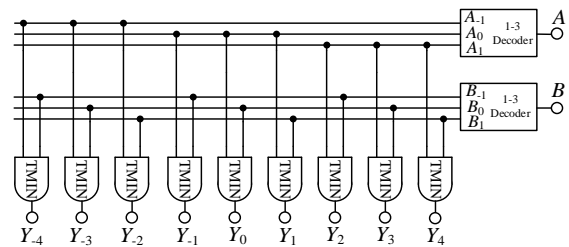

Figure 11 Schematic diagram of balanced ternary 2-9 decoder

Taking the input combination as $A=-1$ and $B=1$ as an example, at this time, the output of the 1-3 decoder only has $A_{-1}$ and $B_1$ as logic "1", $A_0$, $A_1$, $B_{-1}$ and $B_0$ are logic "-1", therefore, after the TMINs, only $Y_{-2}$ is logic "1", and the rest of the output terminals are logic "-1". Figure 12 shows the corresponding simulation results for 2-9 decoder.

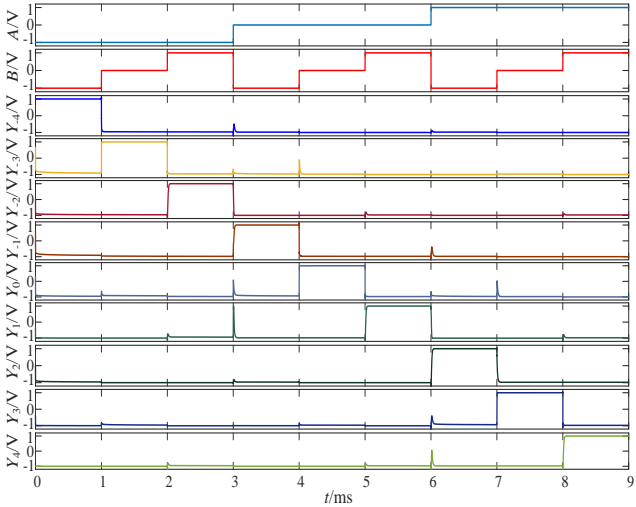

Figure 12 Simulation results of balanced ternary 2-9 decoder

### 4.3 Design of Functional Balanced Ternary Combinational Logic Circuits Based on Decoder

In this section, the balanced ternary functional combinational logic circuits based on the 1-3 decoder are designed, including half adder, multiplier and numerical comparator.

Table 7 shows the truth table of these three functional combinational logic circuits, where HA-S (Half Adder - Sum) and HA-C (Half Adder - Carry) are the sum and carry of the half adder, respectively. MUL (Multiplier) is the output of the multiplier; MLE (More, Less, Equality) is the output of the numerical comparator, the logic "-1" means "$A<B$", the logic "0" means "$A=B$", the logic "1" means "$A>B$".

Table 7 Truth table of balanced ternary half adder, multiplier and numerical comparator

| $A$ | $B$ | HA-S | HA-C | MUL | MLE |
|---|---|---|---|---|---|
| -1 | -1 | 1 | -1 | 1 | 0 |
| -1 | 0 | -1 | 0 | 0 | -1 |
| -1 | 1 | 0 | 0 | -1 | -1 |
| 0 | -1 | -1 | 0 | 0 | 1 |
| 0 | 0 | 0 | 0 | 0 | 0 |
| 0 | 1 | 1 | 0 | 0 | -1 |
| 1 | -1 | 0 | 0 | -1 | 1 |
| 1 | 0 | 1 | 0 | 0 | 1 |
| 1 | 1 | -1 | 1 | 1 | 0 |

| $A$\$B$ | -1 | 0 | 1 | -1 | 0 | 1 |
|---|---|---|---|---|---|---|
| -1 | 1 | -1 | 0 | -1 | 0 | 0 |
| 0 | -1 | 0 | 1 | 0 | 0 | 0 |
| 1 | 0 | 1 | -1 | 0 | 0 | 1 |
| | | S | | | C | |

Figure 13 Karnaugh map of balanced ternary half adder

The balanced ternary decoder can realize the function of decoding one ternary signal into three binary signals. Therefore, after the two input signals $A$ and $B$ pass through the two ternary 1-3 decoders, the outputs $A_{-1}$, $A_0$, $A_1$ and $B_{-1}$, $B_0$, $B_1$ can be obtained respectively, and each output has only two states that logic "1" and logic "-1". Take input $A=B$="0" as an example, at this time, $A_{-1}= A_1= B_{-1}= B_1$="-1", $A_0= B_0$="1". Using the working characteristics of the balanced ternary 1-3 decoder, combined with the Karnaugh map of the balanced ternary half adder shown in Figure 13, the logical expressions of the sum $S$ and the carry $C$ can be deduced as follows:

$$S=0\cdot(A_{-1}B_1+ A_0B_0+ A_1B_{-1}) + (A_{-1}B_{-1}+ A_0B_1+ A_1B_0) \quad (2)$$

$$C=0\cdot(A_{-1}B_0+A_{-1}B_1+ A_0B_{-1}+ A_0B_0 + A_0B_1+A_1B_{-1}+ A_1B_0) +(A_1B_1) \quad (3)$$

Among them, "•" represents the minimum operation (some "•" are omitted, such as $A_{-1}B_1$), and "+" represents the maximum operation. The circuit structure of the designed balanced ternary half adder is shown in Figure 14.

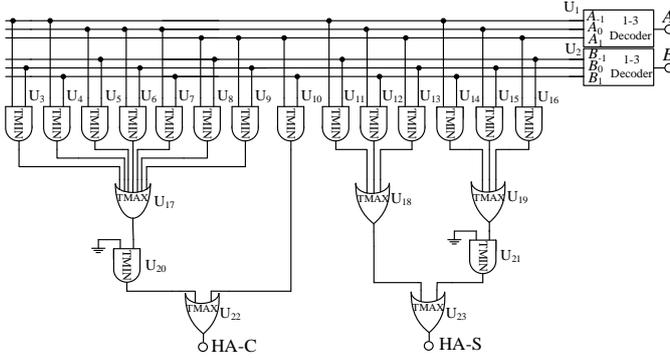

Figure 14 Balanced ternary half adder circuit designed based on the decoder method

Compared with the positive ternary multiplier, the multiplication operation of the balanced ternary does not generate carry, so it only needs two input terminals and one output terminal. Its logical Karnaugh map is shown in Figure 15. Combined with the working characteristics of 1-3 decoder, its logical expression can be deduced:

$$MUL = 0 \cdot (A_{-1}B_0 + A_0B_{-1} + A_0B_0 + A_0B_1 + A_1B_0) + (A_{-1}B_{-1} + A_1B_1) \quad (4)$$

| A\B | -1 | 0 | 1 |
|---|---|---|---|
| -1 | 1 | 0 | -1 |
| 0 | 0 | 0 | 0 |
| 1 | -1 | 0 | 1 |

MUL

Figure 15 Karnaugh map of balanced ternary multiplier

Where $A_{-1}$, $A_0$, $A_1$, $B_{-1}$, $B_0$, $B_1$ are the outputs of two 1-3 decoders respectively. According to formula (4), only two 1-3 decoders, eight 2-input TMINs, two 2-input TMAXs, and one 5-input TMAX are required to complete the design of balanced ternary multiplier, as shown in Figure 16.

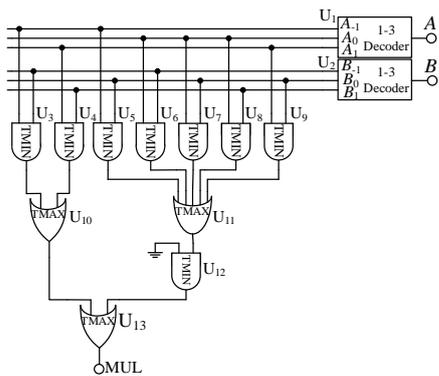

Figure 16 Balanced ternary multiplier circuit designed based on the decoder method

For example, when $A=1$, $B=1$, $A_1$ and $B_1$ are both logic "1" at this time, $A_{-1}$, $A_0$, $B_{-1}$, $B_0$ are logic "-1", therefore, only the output terminal of $U_4$ is logic "1", and other TMINs $U_3$, $U_5$-$U_9$ all output logic "-1". At this time, the output of $U_{10}$ is logic "1", the output of $U_{11}$ is logic "-1", and after performing the minimum operation with logic "0", the output of $U_{12}$ is logic "-1", so the TMAX $U_{12}$ finally outputs logic "1".

The balanced ternary numerical comparator can also be implemented by decoder-based method. For example, we proposed a design scheme of a positive ternary numerical comparator in [23]. The circuit has three output terminals, which are used to represent $A>B$, $A=B$ and $A<B$, the effective signal of the output terminal is high level signal, which requires 10 transistors and 41 memristors. This paper proposes a more optimal design scheme, which can reduce the number of components, requiring only 10 transistors and 32 memristors. Because ternary logic has three different logic states, only one output can be used to represent the results of three numerical comparisons.

Similarly, according to the Karnaugh map of the ternary numerical comparator shown in Figure 17 and the working characteristics of the 1-3 decoder, its logical expression can be deduced as follows:

$$MLE = 0 \cdot (A_{-1}B_{-1} + A_0B_0 + A_1B_1) + (A_0B_{-1} + A_1B_{-1} + A_1B_0) \quad (5)$$

It can be seen that only two 1-3 decoders, seven 2-input TMIN gates, one 2-input TMAX and two 3-input TMAX gates are required. As shown in Figure 18, it is the circuit structure of the balanced ternary numerical comparator designed based on the decoder method.

| A\B | -1 | 0 | 1 |
|---|---|---|---|
| -1 | 0 | -1 | -1 |
| 0 | 1 | 0 | -1 |
| 1 | 1 | 1 | 0 |

MLE

Figure 17 Karnaugh Map of Balanced Ternary Numerical Comparator

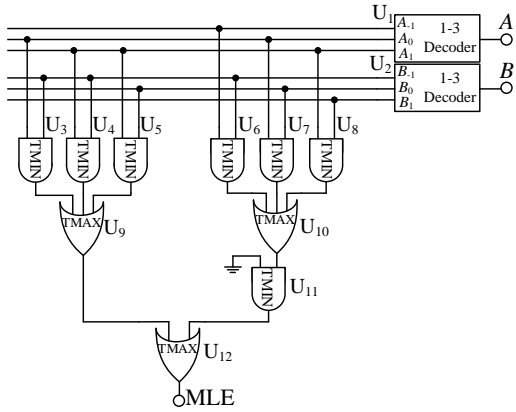

Figure 18 Balanced ternary numerical comparator circuit designed based on the decoder method

In order to verify the validity of the above circuits, each circuit was simulated in LTSpice with the simulation results of the balanced ternary half adder, multiplier and numerical comparator based on the decoder method shown in Figure 19. The results confirm the validity of the circuit designs.

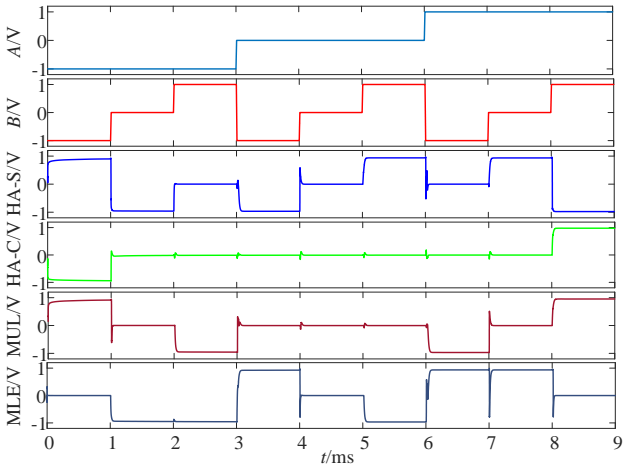

Figure 19 Simulation results of balanced ternary half adder, multiplier and numerical comparator designed based on the decoder method

## 5. Design of the Balanced Ternary Combinational Logic Circuits Based on Multiplexer

In digital logic circuits, it is sometimes necessary to select a particular output signal from multiple input signals, which requires a multiplexer circuit. In addition, multiplexers play an important role in the design of any combinational logic function circuit. In this section, we proposed a design for a balanced ternary multiplexer at first, and then a balanced ternary half adder, multiplier, and numerical comparator are designed on this basis.

Table 8 The input-output relationship of balanced ternary multiplexer

| S | OUT |
|---|---|
| -1 | $I_{-1}$ |
| 0 | $I_0$ |
| 1 | $I_1$ |

### 5.1 Balanced Ternary Multiplexer

The logic function of the multiplexer is to select different input terminals according to the selection signals. Table 8 shows the input-output relationship of the designed balanced ternary multiplexer.

$S$ is the selection signal, $I_{-1}$, $I_0$, $I_1$ are three input signals, and OUT is the output terminal. When $S$ is logic "-1", the output of the circuit is equal to the logic value of $I_{-1}$, when $S$ is logic "0", OUT is equal to the logic value of $I_0$, when $S$ is logic "1", OUT is the state of $I_1$.

The multiplexer designed in this paper includes a balanced ternary 1-3 decoder, three 2-input TMIN gates and a 3-input TMAX gate. Its circuit structure is shown in Figure 20.

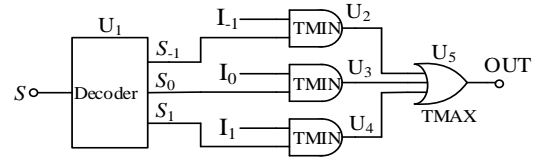

Figure 20 Circuit structure of balanced ternary multiplexer

When the selection signal $S$ is logic "-1", the output terminal $S_{-1}$ of the decoder $U_1$ in Figure 20 is logic "1", and the outputs of the $S_0$ and $S_1$ terminals are logic "-1". At this time, the state of the output of $U_2$ is equal to the state of the input of $I_{-1}$, and the outputs of $U_3$ and $U_4$ are both logic "-1". Therefore, the three inputs of the $U_5$ are the state of $I_{-1}$, logic "-1", logic "-1", so the output is the state of $I_{-1}$. When the selection signal $S$ is logic "0", the output terminal $S_0$ of $U_1$ is logic "1". At this time, the state of the output terminal of $U_3$ is equal to the state of $I_0$, and the output terminals of $U_2$ and $U_4$ are both logic "-1", so the final output of the circuit is the state of $I_0$. When the selection signal $S$ is logic "1", $S_1$ is logic "1". At this time, the state of the output terminal of $U_4$ is equal to the state of the input terminal of $I_1$, and the output terminals of $U_2$ and $U_3$ are both logic "-1". Therefore, the output terminal of $U_5$ is the state of $I_1$. Figure 21 shows the LTSpice simulation results for the balanced ternary multiplexer.

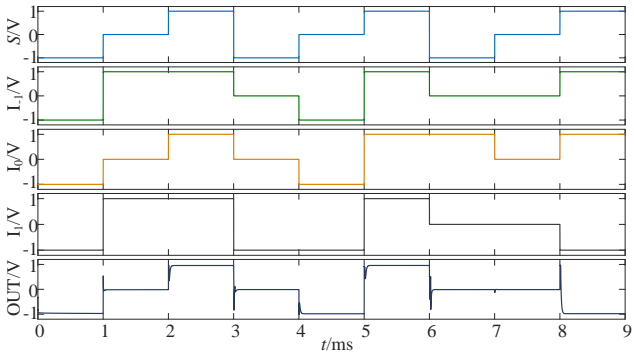

Figure 21 Simulation results of balanced ternary multiplexer

## 5.2 Design of Functional Balanced Ternary Combinational Logic Circuits Based on Multiplexer

When using this method to design balanced ternary logic circuits, the input signals are used as the selection signals of the two-stage multiplexers and the three logic levels are used as the inputs of the first-stage multiplexers. The outputs of the first-stage multiplexers are used as the inputs of the second-stage multiplexers, and the outputs of the second-stage multiplexers are the output terminals of the entire circuit. Figure 22 shows the logical structure diagram of the balanced ternary half adder designed based on the multiplexer method.

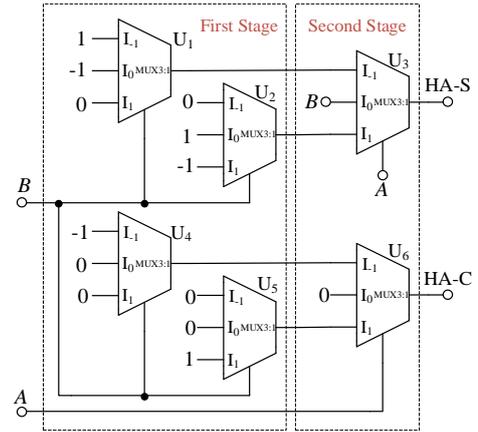

Figure 22 Balanced ternary half adder designed based on the multiplexer method

For the half adder circuit shown in Figure 22, $U_1$, $U_2$, $U_4$, and $U_5$ are the first-stage multiplexers, and their selection signals are all input signal $B$, and $U_3$ and $U_6$ are the second-stage multiplexers, their selection signals are both another input $A$. Among them, $U_1$-$U_3$ constitute the "sum" output circuit of the half adder and the multiplexers $U_4$-$U_6$ constitute the "carry" output circuit. For example, when $A$ is logic "1" and $B$ is logic "0", the $I_0$ input terminals of $U_2$ and $U_5$ are "selected", and the $I_1$ input terminals of $U_3$ and $U_6$ are "selected". Therefore, the output sum $S$ is logic "1", and the carry $C$ is logic "0".

A balanced ternary multiplier can also be designed and implemented based on two multiplexers. According to the truth table of the balanced ternary multiplier given in Table 7, when one of the inputs $A$ is logic "-1" and $B$ takes logic "-1", "0", and "1", the output is logic "1", "0", "-1" respectively. Therefore, the above logic relationship can be converted by a multiplexer. And when $A$ is logic "0", no matter what value $B$ takes, the output terminal is logic "0". When $A$ is a logic "1", the state of the output is the same as the input $B$. As shown in Figure 23, it is the circuit structure of the balanced ternary multiplier designed based on the multiplexer method.

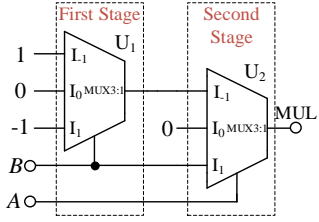

Figure 23 Balanced ternary multiplier designed based on the multiplexer method

Similarly, the balanced ternary numerical comparator can also be implemented by using multiplexer method, specifically, four multiplexers are required. Among them, the input signal *A* is used as the selection signal of the second-stage multiplexer $U_4$, and the three multiplexers $U_1$-$U_3$ are the first-stage multiplexers, as shown in Figure 24.

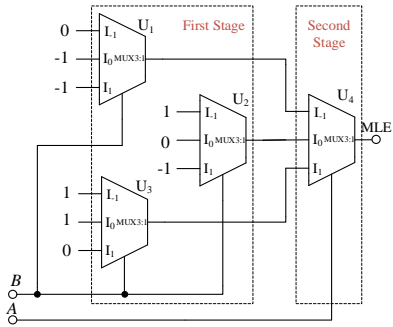

Figure 24 Balanced ternary numerical comparator designed based on the multiplexer method

Figure 25 shows the LTSpice simulation results for the balanced ternary half adder, multiplier, and numerical comparator designed based on the multiplexer method.

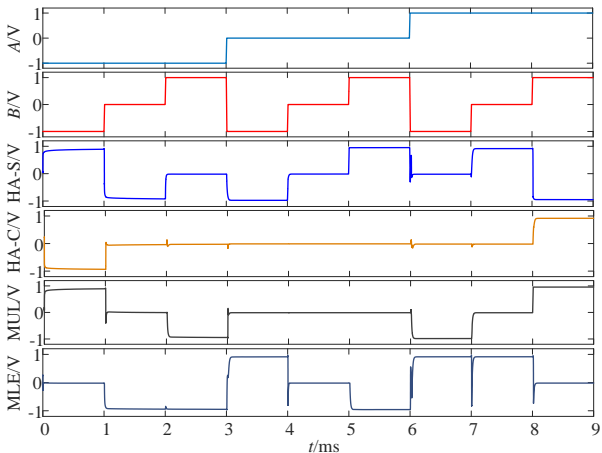

Figure 25 Simulation results of balanced ternary half adder, multiplier and numerical comparator designed based on the multiplexer method

## 6. Experimental Results

To validate the effectiveness of the proposed balanced ternary digital logic circuits, we conducted two experiments utilizing a balanced ternary 3-1 encoder and a half adder as illustrative examples.

To achieve the physical implementation of the designed 3-1 encoder circuit, this research involved the fabrication of bipolar metal-oxide-based memristor with reliable resistive switching behavior[24]. The device exhibits bipolar memristive switching characteristics, demonstrating its ability to transition between high resistance states (HRS) and low resistance states (LRS). As shown in Figure 26, the device exhibits repeatable SET and RESET operations when subjected to multiple sweeps of positive and negative voltages. During the SET process, the voltage varies between 0.4 to 0.7V, indicating the device's transition from HRS to LRS, as depicted by curve 1 and curve 2. To mitigate the risk of breakdown resulting from high currents during the SET process, a compliance current preset of 0.01mA has been employed. Conversely, during the RESET process, the voltage ranges varies from -0.3 to -0.4V, and the transition from LRS back to HRS is depicted by curves 3 and curve 4.

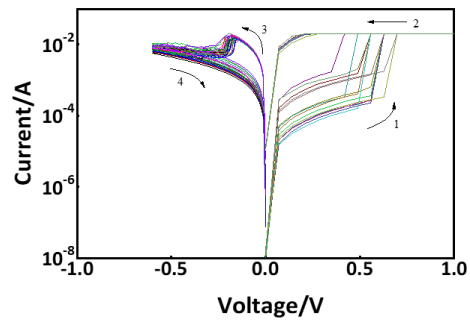

Figure 26 Voltage- current characteristics of the bipolar memristor

Non-volatility is a crucial requirement for storage-computing integrated circuits, as it empowers them to exhibit low-power characteristics. Hence, we conducted additional verification of the non-volatile properties inherent in the designed components. As illustrated in Figure 27, we applied a continuous low-level voltage of 0.1V at room temperature to measure the resistance value. Both the high-resistance and low-resistance states endured for a minimum of $10^4$ s. These experimental findings serve as compelling evidence of the circuit's endurance.

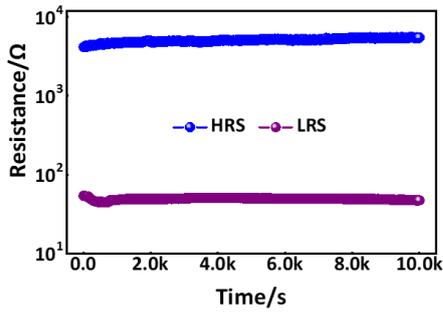

Figure 27 Room-temperature endurance.

Expanding upon this groundwork, we employed a specific configuration of connections involving TMIN gates, TMAX gates, and inverters. For the TMIN gates, we utilized two memristors, with their top electrodes connected as the output terminal and their bottom electrodes linked as the two input signal terminals. In contrast, the three-input TMAX gate employed three memristors interconnected at their bottom electrodes to serve as the output, while their top electrodes functioned as the three input signal terminals. For a comprehensive understanding of the circuit's structure and operational principles, the reader is referred to previous work[19]. These experimental results conclusively demonstrate the circuit's successful implementation as a 3-1 encoder, as visually represented in Figure 28.

In the context of the half adder circuit, it is worth noting that while the overall connection configuration of the circuit may appear complex, it is important to highlight that a significant portion of the sub-module circuits maintain a consistent structure. Hence, in order to accommodate the limited availability of devices, each stage was tested independently. As for the balanced ternary half adder in Figure 14, the decoder stage was first tested, followed by the TMIN stage, followed by the output generation with both TMIN and TMAX gates. As for the half adder in Figure 22, each multiplexer used was individually tested.

The experimental results of the balanced ternary half adder designed based on the decoder and multiplexer methods are shown in Figure 29 and Figure 30, respectively. It can be seen from the figures that the experimental results are mostly consistent with the SPICE simulation results.

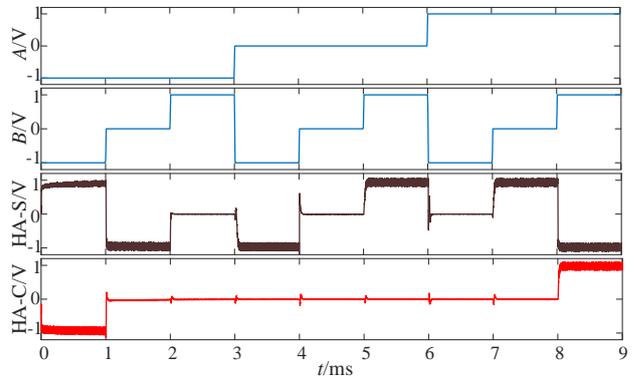

Figure 29 Experimental results of the balanced ternary adder designed based on the decoder-based method

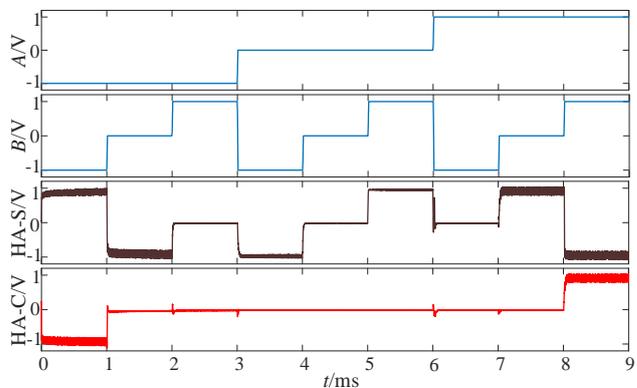

Figure 30 Experimental results of the balanced ternary adder designed based on the multiplexer-based method

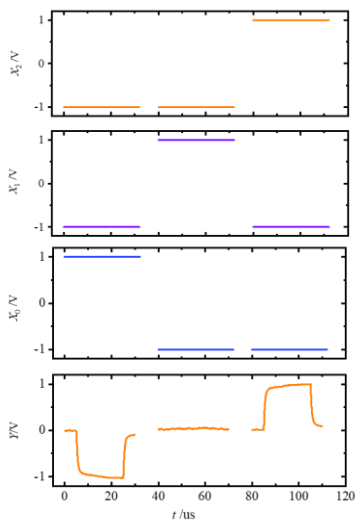

Figure 28 The experiment results of balanced ternary 3-1 encoder

## 7. Comparison and Analysis

For the balanced ternary functional combinational logic circuits designed by the decoder-based method and multiplexer-based method, we have carried out comparison and analysis to demonstrate the advantages and disadvantages of the two methods. Table 9 presents the quantity statistics of the components in the circuits designed above. It is worth noting that the multiplexer is designed based on the 1-3 decoder, therefore in the combinational logic circuits designed based on the multiplexer method, the 1-3 decoders can be reused for multiplexers with the same selection signal, which can further reduce the component consumption in the circuits.

Table 9 Number of components in the circuits designed in this paper

| Method | Components | | |
|---|---|---|---|
| | THA | MUL | MLE |
| Decoder-Based | 10T59M | 10T35M | 10T32M |
| Multiplexer-Based | 10T64M | 10T28M | 10T46M |

[1] T: Transistor. [2] M: Memristor.

According to the statistical results shown in Table 9, the number of components used in the balanced ternary multiplier circuit designed based on the multiplexer method is less than that based on the decoder method, while for the half adder and numerical comparator circuits, the decoder-based approach uses a relatively small number of components. From this it follows that the decoder-based method has an advantage in terms of components numbers.

To further investigate the practical potential of the two methods, we present the power consumption estimates in Table 10. By taking the sum of the voltage-current product through each component, we can get static power dissipation. Average power dissipation is the average value of the static power dissipation for every input combination. Dynamic power consumption is the absolute value of the difference between the maximum instantaneous power consumption and the average power consumption, all of them can be obtained through the SPICE simulations. As shown in Table 10, the average power consumption of the half adder using decoder-based method is reduced by 99.77%. And in terms of dynamic power consumption, the gap between the two method is even more obvious; The half-adder, multiplier and numerical comparator designed based on the multiplexer method are 9.2 times, 6.65 times and 2.23 times that of the decoder-based method respectively. Therefore, the decoder-based approach has a significant advantage in terms of circuit power consumption.

In summary, the decoder-based method has obvious advantages both in terms of component numbers and power consumption, but the multiplexer-based method has the advantages of being based on a simple operating principle and ease of implementation. If multiplexer circuits with lower power consumption could be designed, they would definitely be more widely used.

Table 10 Power consumption statistics of designed circuits by the two methods

| Method | Avg. Power(uW) | | | Static Power(uW) | | | Dynamic Power(mW) | | |
|---|---|---|---|---|---|---|---|---|---|
| | THA | MUL | MLE | THA | MUL | MLE | THA | MUL | MLE |
| Decoder-Based | 0.17 | 0.18 | 0.18 | 3.01 [1&1] | 2.00 [-1&-1] | 2.13 [1&-1] | 0.55 | 0.69 | 0.73 |
| Multiplexer-Based | 72.65 | 72.84 | 0.56 | 201 [-1&1] | 181 [0&1] | 1.51 [0&-1] | 5.06 | 4.59 | 1.63 |

# 8. Conclusion

This paper presented the design methods for memristor-based balanced ternary digital logic circuits. Firstly, balanced ternary basic logic gates TMIN, TMAX and TI were designed based on the binary characteristics of memristor and the switching characteristics of MOSFETs. Balanced ternary encoder, decoder and multiplexer circuits were then designed on this basis and were shown to operate with fewer components than previous designs. Finally, functional combinational logic circuits, including a balanced ternary half adder, multiplier and numerical comparator were designed using two methods: a decoder-based method and a multiplexer-based method. The above circuits were verified and compared by LTSpice simulations, providing new insight into the advantages and disadvantages of the different approaches and a strong foundation for the further development of balanced ternary digital logic circuits.